# Multiscale Graph Neural Networks for Protein Residue Contact Map Prediction

Kuang Liu,[a] Rajiv K. Kalia,[a] Xinlian Liu,[b] Aiichiro Nakano,[a] Ken-ichi Nomura,[a] Priya Vashishta,[a] Rafael Zamora-Resendiz[c]

[a] Collaboratory for Advanced Computing and Simulations, Department of Computer Science, Department of Physics & Astronomy, Department of Chemical Engineering & Materials Science, Department of Quantitative & Computational Biology, University of Southern California, Los Angeles, CA 90089, USA
[b] Department of Computer Science, Hood College, Frederick, MD 21701, USA
[c] Lawrence Berkeley National Laboratory, Berkeley, CA 94720, USA
(liukuang, rkalia, anakano, knomura, priyav)@usc.edu, liu@hood.edu, rzamoraresendiz@lbl.gov

## Abstract

Machine learning (ML) is revolutionizing protein structural analysis, including an important subproblem of predicting protein residue contact maps, *i.e.*, which amino-acid residues are in close spatial proximity given the amino-acid sequence of a protein. Despite recent progresses in ML-based protein contact prediction, predicting contacts with a wide range of distances (commonly classified into short-, medium- and long-range contacts) remains a challenge. Here, we propose a multiscale graph neural network (GNN) based approach taking a cue from multiscale physics simulations, in which a standard pipeline involving a recurrent neural network (RNN) is augmented with three GNNs to refine predictive capability for short-, medium- and long-range residue contacts, respectively. Test results on the ProteinNet dataset show improved accuracy for contacts of all ranges using the proposed multiscale RNN+GNN approach over the conventional approach, including the most challenging case of long-range contact prediction.

## I. Introduction

The three-dimensional (3D) structure of a protein reveals crucial information about how it interacts with other proteins to carry out fundamental biological functions. Proteins are linear chains of amino acids that fold into specific 3D conformations as a result of the physical properties of the amino acid sequence. The structure, in turn, determines the wide range of protein functions. Thus, understanding the complexity of protein folding is vital for studying the mechanisms of these molecular machines in health and disease, and for development of new drugs. Various machine learning (ML) techniques have been applied successfully to protein structure analysis in the past [1, 2]. In previous works, for example, we explored fast atomistic learning based on a 2D convolutional neural network (CNN), through dimension mapping using space-filling curves [3]. We have also demonstrated that a novel spatial model built with a graph convolution network (GCNN) can be used effectively to produce interpretable structural classification [4]. ML models such as neural networks have long been applied to predict 1D structural features such as backbone torsion angles, secondary structure and solvent accessibility of amino-acid residues. The focus of ML applications has since shifted to 2D representation of 3D structures such as residue-residue contact maps [5] and inter-residue distance matrices. Recognizing that contact maps are similar to 2D images — whose classification and interpretation have been among the most successful applications of deep learning (DL) approaches — the community has begun to apply DL to recognize patterns in the sequences and structures of proteins in the protein data bank (PDB) [6]. CNNs have demonstrated excellent performance in image analysis tasks, making them a natural choice for the prediction of protein contact maps. The question of how best to encode information about the target protein for input to the neural network is an active research topic. By analogy, color images are often encoded as three matrices of real numbers, *i.e.*, the intensities of red, green and blue color channels for all image pixels. Methods such as DeepContact [7] and RaptorX-Contact [8] use input features consisting of $N \times N$ (where $N$ is the number of amino acids in the sequence of the target protein) residue-residue coupling matrices derived from covariation analyses of the target protein, augmented by predictions of local sequence features. In the DeepCov [9] and TripletRes [10] approaches, more information in the target protein multiple-sequence alignment is provided to the network, in the form of 400 different $N \times N$ feature matrices, each corresponding to a defined pair of amino acids, with the value at position $(i, j)$ in a given matrix being either the pair frequency or the covariance for the given amino acid pair at alignment positions $i$ and $j$. Then, CNN integrates this massive set of features to identify spatial contacts, training by large sets of proteins of known structure and their associated contact maps and multiple sequence alignments. The importance of incorporating ML in template-free modeling has been highlighted by top-performing CASP13 structure prediction methods, all of which rely on deep convolutional neural networks for predicting residue contacts or distances, predicting backbone torsion angles and ranking the final models.

Approaches to the protein residue contact map prediction problem include support vector machine (SVM) [11], CNN [9], recurrent neural network (RNN) + CNN [12], ResNet, VGG and other proven architectures. Most of these approaches require heavy feature engineering. Apart from the amino acids sequence, they used additional engineered features such as amino-acid pair frequency [9], covariation scores [9, 13], and position-specific scoring matrix (PSSM). Such heavy feature engineering often leads to poor ability of transfer learning across relevant protein folding tasks. Bepler *et al.* [12] demonstrated a promise of transfer learning, *i.e.*, ability to transfer knowledge between structurally related proteins, through representation learning, which we will follow here. In this work, we introduce graph neural network (GNN) to the conventional RNN-based pipeline [12] so as to better capture spatial correlations. Furthermore, we propose a multiscale GNN approach, in which short-, medium- and long-range spatial correlations are refined by dedicated respective GNNs after coarse learning.

## II. Contact Map Prediction through Representation Learning

### *Contact Map Prediction Problem*

A protein contact map represents pairwise amino acid distances, where each pair of input amino acids from sequence is mapped to a label $\in \{0,1\}$, which denotes whether the amino acids are "in contact" (1, *i.e.*, within a cutoff distance of 8 Å) or not (0); see Fig. 1. Accurate contact maps provide powerful global information, *e.g.*, they facilitate the understanding of the complex dynamical behavior of proteins or other biomolecules [14] and robust modeling of full 3D protein structure [15]. Specifically, medium- and long-range contacts, which may be as few as twelve sequence positions apart, or as many as hundreds apart, are particularly important for 3D structure modeling. However, existing approaches [9, 11, 12] require heavy feature engineering and suffer from poor ability of transfer learning across relevant protein folding tasks.

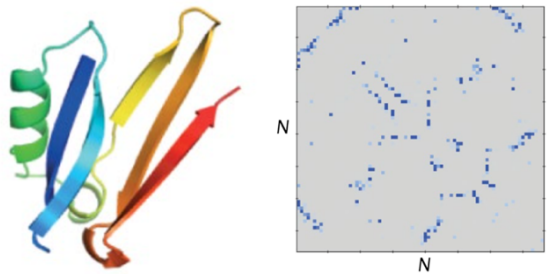

**Fig. 1.** (Left) An example of 3D protein structure. (Right) An example of residue-residue contact map.

### *Proposed Approach*

Here, we address these problems with the proved strength of representation learning. Specifically, we use a type of RNN, *i.e.*, bidirectional long short-term memory (LSTM) embedding model, mapping sequences of amino acids to sequences of vector representations, such that residues occurring in similar structural contexts will be close in embedding space (Fig. 2). How to induce the residue-residue contact from the vector representations is still a challenge, because these vectors have few position-level correspondences between residues. We solve this problem by introducing GNN. The role of GNN is, *via* its powerful capability of structural learning, to infer the pair relation of residues from the intermediate vector representations.

Graph-based data in general can be represented as $\boldsymbol{G} = (\boldsymbol{V}, \boldsymbol{E})$, where $\boldsymbol{V}$ is the set of nodes and $\boldsymbol{E}$ is the set of edges. Each edge $e_{uv} \in \boldsymbol{E}$ is a connection between nodes $u$ and $v$. If $\boldsymbol{G}$ is directed, we have $e_{uv} \not\equiv e_{vu}$; if $\boldsymbol{G}$ is undirected, instead $e_{uv} \equiv e_{vu}$. Here, we deal with undirected graphs, but it is trivial to modify such a model to process other directed graph data. In protein residue contact graphs, the nodes are amino-acid residues and the edges are their spatial proximity within 8 Å.

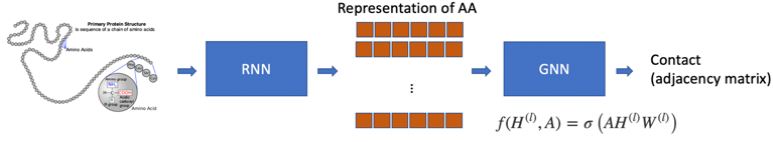

**Fig. 2.** Procedures of the leaning task. AA: amino acid.

The goal of GNN is to learn low-dimensional representation of graphs from the connectivity structure and input features of nodes and edges. The forward pass of GNN has two steps, *i.e.*, message passing and node-state updating. The architecture is summarized by the following recurrence relations, where *t* denotes the iteration count:

$$m_v^{t+1} = \sum_{w \in N(v)} M_t(h_v^t, h_w^t, e_{vw}) \tag{1}$$

$$h_v^{t+1} = U_t(h_v^t, m_v^{t+1}) \tag{2}$$

where $N(v)$ denotes the neighbors of node $v$ in graph $\boldsymbol{G}$. The message function $M_t$ takes node state $h_v^t$ and edge state $e_{vw}$ as inputs and produces message $m_v^{t+1}$, which can be considered as a collection of feature information from the neighbors of $v$. The node states are then updated by function $U_t$ based on the previous state and the message. The initial states $h_v^0$ are set to be the input features of amino acids. Here, we use normalized adjacency matrix $\tilde{A}$ of the graph coupled with some other features as the edge state. These two steps are repeated for a total of $T$ times in order to gather information from distant neighbors, and the node states are updated accordingly. GNN can be regarded as a layer-wise model that propagates messages over the edges and update the states of nodes in the previous layer. Thus, $T$ can be considered to be the number of layers in this model.

The exact form of message function is

$$m_v^{t+1} = A_v \boldsymbol{W^t}[h_1^t \; ... \; h_v^t] + \boldsymbol{b} \tag{3}$$

where $\boldsymbol{W^t}$ are weights of GNN and $\boldsymbol{b}$ denotes bias. We use gated recurrent units as the update function:

$$z_v^t = \sigma(\boldsymbol{W^z} m_v^t + \boldsymbol{U^z} h_v^{t-1}) \tag{4}$$

$$r_v^t = \sigma(\boldsymbol{W^r} m_v^t + \boldsymbol{U^r} h_v^{t-1}) \tag{5}$$

$$\widetilde{h_v^t} = \tanh(\boldsymbol{W} m_v^t + \boldsymbol{U}(r_v^t \odot h_v^{t-1})) \tag{6}$$

$$h_v^t = (1 - z_v^t) \odot h_v^{t-1} + z_v^t \odot \widetilde{h_v^t} \tag{7}$$

where $\odot$ denotes element-wise matrix multiplication and $\sigma(\cdot)$ is sigmoid function for nonlinear activation.

### *Multiscale RNN+GNN Model*

Figure 3 shows a schematic of the proposed multiscale RNN+GNN model, which improves upon existing RNN+CNN models for protein residue contact map prediction [13].

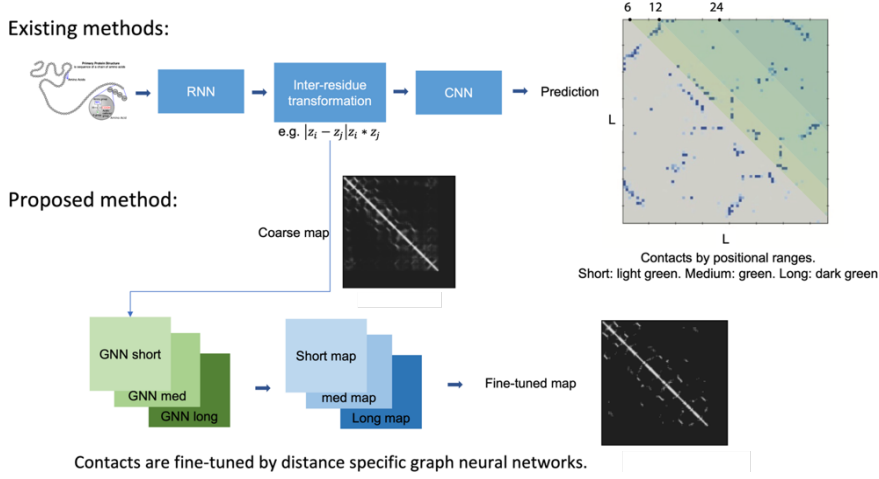

**Fig. 3.** The proposed multiscale GNN approach on top of the existing RNN-CNN pipeline.

First, the RNN unit works as encoder that takes a sequence of amino acids representing a protein and transforms it into vector representations of the same length. To align with the biological process of protein folding, we employ bidirectional LSTM as encoder to absorb the representation from the neighboring amino acids, thus the resultant embeddings contain hidden feature from both sides.

In order to produce contacts from the sequence of embeddings, we define a pairwise feature tensor,

$$v_{ij} = [z_i - z_j \,||\, z_i \odot z_j], \tag{8}$$

of size $L \times L \times 2D$, where $D$ is the dimension of the embedding vector and $L$ is the length of protein, $||$ is concatenation, and $\odot$ is element-wise product. This featurization is symmetric and has extensive applications in natural language processing (NLP) models [16]. This 3D feature is then transformed though the proposed GNN module. To have higher granularity of the contact predictions with regard to the positional ranges, particularly short-, medium- and long-range contacts, we have designed three range-based GNN blocks, where in each layer $t$, the edge feature, $(E_s^t, E_m^t, E_l^t)$, and node feature, $(H_s^t, H_m^t, H_l^t)$, are updated respectively as

$$e_{ij}^{t+1} = \alpha[W_1(\alpha(W_2 e_{ij}^t) \,||\, \alpha(W_3 h_i^t || h_j^t))] \tag{9}$$

$$h_i^{t+1} = \alpha[W_4(\alpha(W_5 h_i^t \,||\, \sum_j e_{ij}^{t+1} / N))] \tag{10}$$

where $W_{1\ldots5}$ are learnable weights and $\alpha$ is the rectified linear unit (ReLU) activation. In Eq. (10), $N$ is the number of positional neighbors of an amino acid, which distinguishes the range-based blocks: (i) short-range blocks focus on positional neighbors from 6 to 11, so that there are $N = 12 - 6 = 6$ neighbors; (ii) medium-range blocks for which $N = 24 - 12 = 12$; and (iii) long-range block for which $N = L - 24$. Thus, the new node features are induced by an average of the neighboring edge features.

The output of the final layer of the GNN blocks is a triplet, $\boldsymbol{E}^T = (\boldsymbol{E}_s^T, \boldsymbol{E}_m^T, \boldsymbol{E}_l^T)$, containing edge features of the corresponding range-based blocks. A fully connected layer will merge and convert $\boldsymbol{E}^T$ into the contact map.

It is worth noting here that our multiscale feature averaging is akin to the additive hybridization scheme [17] used in the celebrated multiscale quantum-mechanical (QM)/molecular-mechanical (MM) simulation approach, for which Karplus, Levitt and Warshel shared the Nobel prize in chemistry in 2013 [18]. In the additive hybridization scheme, energy contributions from different spatial ranges are described by appropriate approaches in the respective ranges which are averaged to provide the total energy [17].

## III.  Results and Discussion

To evaluate the proposed model, we use the ProteinNet dataset [19]. ProteinNet is a standardized dataset for machine learning of protein structure which builds on CASP (Critical Assessment of protein Structure Prediction) assessment carrying out blind prediction of recently solved but publicly unavailable protein structures. In our experiment, we specifically chose a subset from CASP12. We implemented all methods in Tensorflow 2.5 and trained on two NVIDIA V100 graphics processing units (GPUs). The GNN module consists of two layers with independent parameters for each short-, medium- and long-range based block.

Figure 4 shows the precision of the proposed multiscale RNN+GNN approach as a function of the training epoch compared with that of the baseline RNN+CNN approach. Here, we employ a commonly used metrics. The term "P@K" signifies precision for the top $K$ contacts, where all predicted contacts are sorted from highest to lowest confidence. Let $L$ be the length of the protein, then "P@L/2", for example, is the precision for the $L/2$ most likely predicted contacts.

In Fig. 4, we observe improved P@L/2 precision by the addition of multiscale GNNs. As is well known, contact prediction with increased spatial ranges are progressively more difficult. The proposed multiscale RNN+GNN approach consistently outperforms the baseline in all ranges, including the hardest case of long-range contact prediction.

In Tables 1-3, we show the results of contact prediction in terms of P@L, P@L/2 and P@L/5. We benchmark the proposed method against the CNN-based approach [12] as baseline. The precisions are collected from the test dataset consisting of 144 protein sequences. The new multiscale RNN+GNN approach consistently improves the prediction precision over the baseline RNN+CNN approach for all K top contacts (K = L, L/2, L/5). While the precision decreases as we move from the short- to medium- to long-ranges as expected, the multiscale RNN+GNN approach maintains its precision advantage over the baseline.

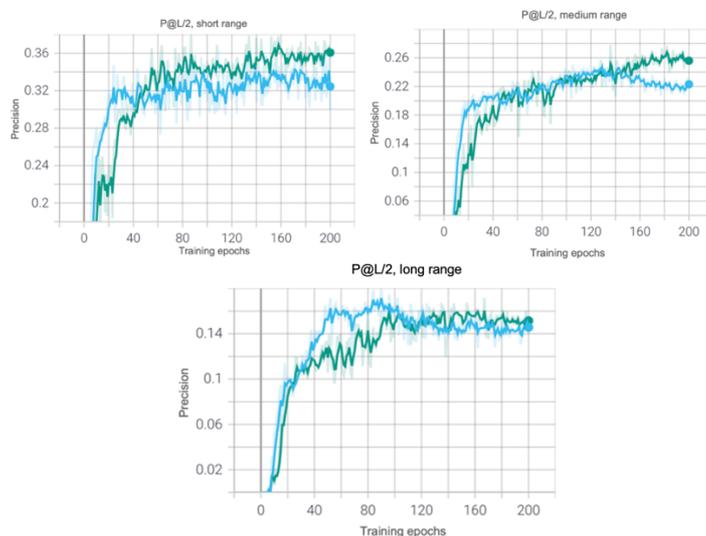

**Fig. 4.** P@L/2 precision of the proposed multiscale RNN+GNN model (green) as a function of the training epoch compared to that of the conventional RNN+CNN model (cyan) for the prediction of (top) short-, (middle) medium- and (bottom) long-range protein residue contacts.

**Table 1.** Short-range contact prediction results.

|  | P@L | P@L/2 | P@L/5 |
|---|---|---|---|
| baseline | 0.319 | 0.299 | 0.344 |
| proposed | 0.360 | 0.355 | 0.370 |

**Table 2.** Medium-range contact prediction results.

|  | P@L | P@L/2 | P@L/5 |
|---|---|---|---|
| baseline | 0.220 | 0.223 | 0.237 |
| proposed | 0.254 | 0.256 | 0.263 |

**Table 3.** Long-range contact prediction results.

|  | P@L | P@L/2 | P@L/5 |
|---|---|---|---|
| baseline | 0.135 | 0.139 | 0.158 |
| proposed | 0.145 | 0.150 | 0.165 |

## IV. Concluding Remarks

In summary, we have proposed a multiscale GNN-based approach taking a cue from the celebrated multiscale physics simulation approach, in which a standard pipeline consisting of RNN and CNN is improved by introducing three GNNs to refine predictive capability for short-, medium- and long-range contacts, respectively. The results show improved accuracy for contacts of all ranges, including the most challenging case of long-range contact prediction. The multiscale GNN approach reflects the inherently multiscale nature of biomolecular and other physical systems, and as such is expected to find broader applications beyond the protein residue contact prediction problem.

As deep learning continues to show promise at modeling the relation between primary and tertiary structure, methods of interpreting learned representations become progressively more important in providing biologically meaningful insights about how proteins work. Previous work has tackled approaches for identifying biologically significant model parameters such as saliency measures along 3D space for volumetric methods like CNNs [4]. However, these approaches are limited in that saliency maps over a volume do not adequately describe the role of interactions between residues or the predictive value of higher-order sub-structures. Work by Zamora-Resendiz *et al*. [3] demonstrated how GCNN architectures learn more biologically relevant representations. Parameter attributions were found to localize at meaningful segments (including secondary structures) for RAS proteins and these sub-structures were found to be characterized in literature on RAS. As we learn more about how to represent physical systems in deep learning frameworks, the "data-agnostic" capabilities of deep learning methods will help in discovering biologically relevant sub-structures given the proper innate priors.

With the continuously growing size of the ProteinNet dataset used in this study, the proposed ML approach is becoming heavily compute bound. To address this challenge, we are currently implementing our model on leadership-scale parallel supercomputers at the Argonne Leadership Computing Facility (ALCF), including the Theta [20] and new Polaris machines. Each computing node of Polaris consists of one AMD EPYC "Milan" central processing unit (CPU) and four NVIDIA A100 GPUs. These leadership-scale implementations will be applied to the largest-available protein datasets. For the massively parallel learning, we employ data parallelism utilizing a distributed ML framework, Horovod [21]. Here, the global batch of input data are split across computing nodes, and the model parameters are updated by aggregating the gradients from the nodes. Up to $O(100)$ nodes, we adopt synchronous training, where the model and gradients are always in sync among all the nodes. For larger-scale training, hybrid synchronous-asynchronous approach will be employed instead for higher scalability (though with slower statistical convergence) [22]. For complex network and hyperparameter tuning, we also use DeepHyper, a scalable automated ML (AutoML) package for the development of deep neural networks [23]. In particular, we utilize two components of the package: (i) neural architecture search (NAS) for automatic search of high-performing deep neural network (DNN) architectures; and (ii) hyperparameter search (HPS) for automatic identification of high-performing DNN hyperparameters. Running such massive ML workflow on leadership-scale parallel supercomputers will likely pose runtime challenges such as fault recovery, for which we will utilize ALCF support for Balsam high performance computing (HPC) workflows and edge services [24]. The multiscale RNN+GNN model is being scaled up to leadership-scale parallel supercomputers.

## V. Acknowledgments

This work was supported by the National Science Foundation, CyberTraining Award OAC 2118061. Some calculations were performed at the Center for Advanced Research Computing (CARC) of the University of Southern California. Scalable parallel implementation is being performed at the Argonne Leadership Computing Facility under the DOE INCITE and Aurora Early Science programs.